\documentclass[12pt]{iopart}
\usepackage{epsfig,amssymb,subfigure,float}
\begin{document} 
\newcommand {\cs}{$\clubsuit$}
\renewcommand{\baselinestretch}{2} 
\title{Boson-fermion demixing in a cloud of lithium atoms in a pancake trap}
\author{Z. Akdeniz$^{1,2}$, P. Vignolo$^1$ and M. 
P. Tosi$^1\footnote{Author to whom any correspondence should be addressed ({\tt tosim@sns.it})}$}
\address{$^1$NEST-INFM and Classe di Scienze, Scuola Normale Superiore,
I-56126, Pisa, Italy\\
$^2$Department of Physics, University of Istanbul, Istanbul, Turkey}

\begin{abstract}
We evaluate the equilibrium state of a mixture of $^7$Li and $^6$Li atoms
with repulsive interactions, confined 
inside a pancake-shaped trap under conditions such that the thickness of 
the bosonic and fermionic clouds is approaching the values of the 
$s$-wave scattering lengths. 
In this regime the effective couplings depend on the axial confinement 
and full demixing can become observable by merely squeezing the trap, without 
enhancing the scattering lengths through recourse to a Feshbach resonance.
\end{abstract}

\vspace{0.5cm}
\maketitle

\section{Introduction}
Dilute boson-fermion mixtures have recently been studied
in several experiments by trapping and cooling gases of mixed
alkali-atom isotopes 
\cite{Schreck2001a,Goldwin2002a, Hadzibabic2002a,Roati2002a,Modugno2002a,Ferlaino2003a}.
The boson-fermion coupling strongly affects the equilibrium properties
of the mixture and can lead to quantum phase
transitions. In particular, it has been shown that 
boson-fermion repulsions in three-dimensional (3D) clouds can induce spatial demixing of
the two components when the interaction energy overcomes the kinetic
and confinement energies \cite{Molmer1998a}. Several configurations
with different topology are possible for a demixed cloud inside a
harmonic trap \cite{Akdeniz2002a}. Although spatial demixing has not yet
been experimentally observed, the experiments of Schreck
{\it et al.}  \cite{Schreck2001a} on a $^6$Li-$^7$Li mixtures inside
an elongated 3D trap appear to be not far from the
onset of a demixed state: an increase 
in the boson-fermion scattering length by a factor five would be 
needed to enter the phase-separated regime \cite{Akdeniz2002b}. 

In this Letter we examine the possibility of attaining phase separation by 
merely varying the trap 
geometry for the $^6$Li-$^7$Li mixture at the 
"natural" values of the scattering lengths. We 
focus on the case of quasi-two-dimensional (Q2D) confinement inside 
a pancake-shaped trap. 
We find that this geometry favours the demixed state through the 
appearance of the axial size of the atomic clouds in their effective 
coupling in the azimuthal plane. We also show that several 
configurations are possible in the plane of the trap at zero temperature, 
but the configuration consisting of a core of bosons surrounded 
by a ring of fermions is the most energetically favourable for a wide 
range of values of system parameters.

\section{The model} \label{method}
The atomic clouds are trapped in pancake-shaped potentials given by
\begin{equation}
\hspace{-2cm}V^{ext}_{b,f}(x,y,z)=m_{b,f}
\omega^2_{x(b,f)}(x^2+\lambda^2_{b,f} y^2)/2+m_{b,f}\omega^2_{z(b,f)}z^2/2
\equiv V_{b,f}({x,y})+U_{b,f}(z)
\end{equation} 
where $m_{b,f}$  are the atomic masses and
$\omega_{z(b,f)}\gg\omega_{x(b,f)}$ the trap frequencies.
We focus on the case where the trap is flat enough that 
the dimensions of the clouds in the axial direction, which are of the 
order of the
axial harmonic-oscillator lengths 
$a_{z(b,f)}=(\hbar/m_{b,f}\omega_{z(b,f)})^{1/2}$, are comparable
to the 3D boson-boson and boson-fermion scattering lengths 
$a_{bb}$ and $a_{bf}$.
In this regime we can study the equilibrium properties of the mixture
at essentially zero temperature ($T\simeq 0.02\,T_F$)
in terms of the particle density 
profiles in the $\{x,y\}$ plane, 
which are $n_{c}({x,y})$ for the Bose-Einstein condensate
and $n_{f}({x,y})$ for the
fermions. The profiles are evaluated
using the Thomas-Fermi approximation for the condensate and the
Hartree-Fock approximation for the fermions.
We shall take $a_{zb}=a_{zf}$ $(=a_z$, say) and only near the end we 
shall discuss the case $\lambda_{b,f}\ne 1$.

The Thomas-Fermi approximation assumes that the number of condensed
bosons is large enough that the kinetic energy term in the
Gross-Pitaevskii equation may be neglected~\cite{Baym1996a}. It yields  
\begin{equation}
n_c({x,y})=[\mu_b-V_b({x,y})-g_{bf}n_f({x,y})]/g_{bb}
\label{zehra1}
\end{equation}
for positive values of the function in the square brackets and zero
otherwise. Here, $\mu_b$ is the chemical potential of the bosons.
This mean-field model is valid when the high-diluteness 
condition $n_ca_{bb}^2\ll 1$ holds and the temperature is outside the critical 
region.
If the conditions $a_z>  a_{bb},a_{bf}$
are fulfilled, the atoms experience collisions in three dimensions
and the coupling constants in Eq. (\ref{zehra1}) can be written in terms 
of the 3D ones as~\cite{Lieb2001}
\begin{equation}
g_{bb}=\frac{g_{bb}^{3D}}
{\sqrt{2\pi} a_z}\;,\;\;\;g_{bf}=\frac{g_{bf}^{3D}}{\sqrt{2\pi} a_z}  
\label{2Dint}
\end{equation} 
with $g_{bb}^{3D}=4\pi\hbar^2a_{bb}/ m_b$, 
$g_{bf}^{3D}=2\pi\hbar^2a_{bf}/m_r$ and
$m_r=m_bm_f/(m_b+m_f)$.

The Hartree-Fock
approximation~\cite{Minguzzi1997a,Amoruso1998a,Vignolo2000b}
treats the fermion cloud as an ideal gas subject to an effective 
external potential, that is 
\begin{equation}
n_{f}({x,y})=\int\frac{d^2p}{(2\pi\hbar)^2}\left\{\exp\left[\left(
\frac{p^2}{2m_{f}}
+V^{eff}({x,y})-\mu_{f}\right)/k_BT\right]+ 1\right\}^{-1},
\label{zehra4}
\end{equation}
where $\mu_f$ is the chemical potential of the fermions and
\begin{equation}
V^{eff}({x,y})=V_{f}({x,y})+g_{bf}n_c({x,y}).
\label{zehra3}
\end{equation}
The fermionic
component has been taken as a dilute spin-polarized gas, for which the
fermion-fermion  $s$-wave scattering processes are inhibited  by
the Pauli principle and $p$-wave scattering is
negligibly small~\cite{Demarco1999b}. 
In the mixed regime the 
Fermi wave number in the azimuthal plane should be smaller than $1/a_{bf}$, 
but this is not
a constraint in the regime of demixing where the boson-fermion
overlap is rapidly dropping. 

The chemical potentials $\mu_{b,f}$  characterize the system in the 
grand-canonical ensemble and are determined by requiring that
the integrals of the densities over the $\{x,y\}$ plane 
should be equal to the  average numbers $N_b$ and $N_f$ of particles.
The presence of a bosonic thermal cloud can be 
treated by a similar Hartree-Fock approximation \cite{Akdeniz2002b}, 
but it has quite negligible effects at the 
temperatures of present interest.
\section{General conditions for phase separation}
\label{tzero}
The effective strength of the atom-atom collisions in the azimuthal 
plane depends in our Q2D 
model on the axial harmonic-oscillator length according to Eq. (\ref{2Dint}). 
We derive and illustrate in this section the consequences for demixing 
at very low temperatures.
	
We preliminarly recall that in the macroscopic limit the phase transition 
is sharp and the overlap between the two species after demixing is 
restricted to the interfacial region, where it is 
governed by the surface kinetic energy (see Ref. \cite{Ao1998a} 
for an example in a 3D model). In the 
case of mesoscopic clouds under confinement the transition is instead 
spread out and several alternative definitions of the location of 
demixing can therefore be given. We discuss below 
three alternative locations, that we denote as partial, dynamical, 
and full demixing. Their definition and the derivation of simple 
analytical expressions in the Q2D model are given in the following.

\subsection{Partial demixing}
\label{partial_sep}

The interaction energy $E_{int}$ between the boson and fermion clouds 
initially grows as the boson-fermion coupling is increased, 
but reaches a maximum and then falls off as the overlap between 
the two clouds diminishes. We locate partial demixing at the maximum of 
$E_{int}$, that we calculate from the density profiles according to 
the expression
\begin{equation}
E_{int}=g_{bf} \int dx\,dy \, n_c(x,y) n_f(x,y).
\end{equation}
In the calculations that we report in this section we have used 
values of system parameters as 
appropriate to the experiments in Paris on the
$^6$Li-$^7$Li mixture \cite{Schreck2001a}, that is 
$N_b =N_f\simeq 10^4$ and
$\omega_{xb}/2\pi=4000$~Hz, $\omega_{xf}/2\pi=3520$~Hz.

\begin{figure}[H]
\centerline{
{\epsfig{file=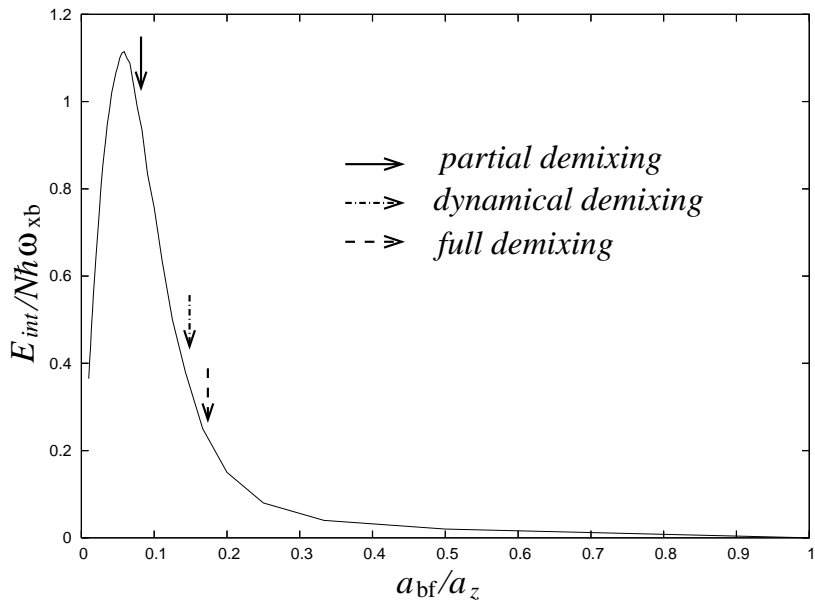,width=0.7\linewidth}}}
\caption{Boson-fermion interaction energy $E_{int}$ in a circular disc
(in units of $N\hbar\omega_{xb}$, with
$N=N_b+N_f$), as a function of $a_{bf}/a_z$ for fixed values of $a_{bb}$ and
$a_{bf}$. The arrows indicate the locations of demixing
estimate from Eqs. (\ref{partially0}), (\ref{dyn0}) 
and (\ref{condition}).}
\label{fig1}
\end{figure}

We first fix the two scattering lengths at their "natural" 
values ($a_{bb} = 5.1\,a_0$ and $a_{bf} = 38\,a_0$, with $a_0$ the Bohr 
radius) and vary the thickness $a_z$ of the trap from $100\,a_{bf}$ down to 
$a_{bf}$. By 
this choice we are increasing both the boson-boson and the 
boson-fermion effective coupling in the azimuthal plane. 
The behaviour of the interaction energy as a function of the ratio 
$a_{bf}/a_z$ at fixed $a_{bb}$ and $a_{bf}$ is shown in 
Fig. \ref{fig1}. It is seen that $E_{int}$ goes through a maximum at 
$a_z \simeq 16\,a_{bf}$.

\begin{figure}[H]
\centerline{
{\epsfig{file=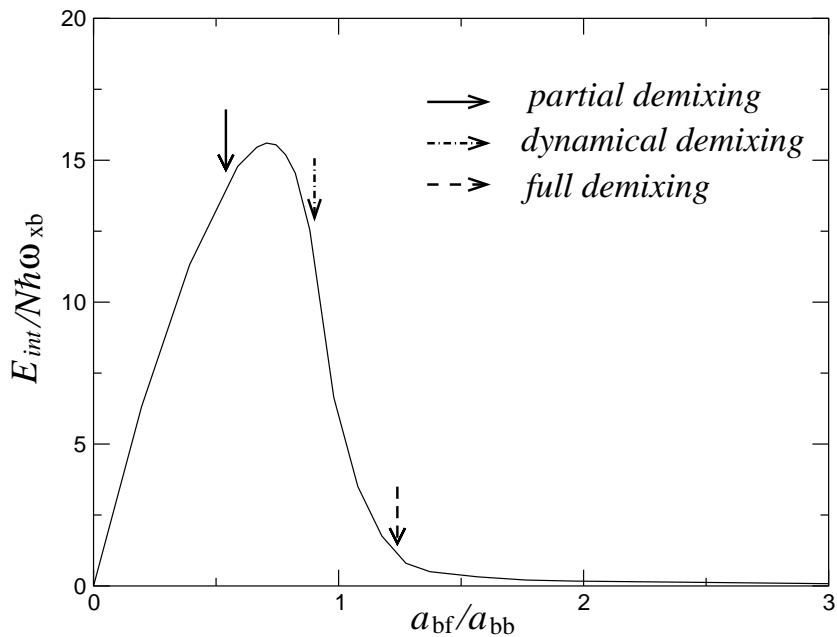,width=0.7\linewidth}}}
\caption{Boson-fermion interaction energy $E_{int}$ in a circular disc
(in units of $N\hbar\omega_{xb}$), as a function of $a_{bf}/a_{bb}$ 
for $a_{bb}=5.1\,a_0$. The arrows indicate the locations of demixing
estimate from Eqs. (\ref{partially0}), (\ref{dyn0}) 
and (\ref{condition}).}
\label{fig2}
\end{figure}

As an alternative and with the aim of a later comparison with 
the results previously 
obtained in the 3D case~\cite{Akdeniz2002a,Akdeniz2002b},
we show in Fig. \ref{fig2} the interaction energy as a function of the ratio 
$a_{bf}/a_{bb}$ at fixed $a_{bb}$.
The axial thickness $a_z$ has been set equal to the largest 
of the two scattering lengths. This condition fixes the limit
of validity of our model~\cite{Tanatar2002a}. 
The maximum of $E_{int}$ lies at $a_{bf}/a_{bb} \simeq 0.8$, much below 
the value $a_{bf}/a_{bb} \simeq 7$ of the ratio 
of the "natural" scattering lengths. In fact, at $a_{bf}/a_{bb} \simeq 7$ 
the two clouds would undergo demixing as the axial trap stifness 
is increased towards a Q2D model.

An approximate analytic formula for the location of partial demixing 
as a function of the system parameters in the Q2D model can be 
obtained from the condition 
$\partial E_{int}/\partial g_{bf}=0$ by 
using the estimate $n_{c,f}\approx N_{b,f}/(2 \pi R_{b,f}^2)$ 
with the values of the cloud radii in the absence of 
boson-fermion interactions,
$R_f=(8 N_f/\lambda_f)^{1/4}a_{xf}$ and $R_b=(16
N_b a_{bb}/(\sqrt{2\pi}a_z\lambda_b))^{1/4}a_{xb}$ 
where $a_{x(b,f)}=(\hbar/m_{b,f}\omega_{x(b,f)})^{1/2}$.
The location of the maximum in the boson-fermion interaction 
energy as a function of $a_{bf}/a_{bb}$ lies at
\begin{equation}
\left.\frac{a_{bf}}{a_{bb}}\right|_{part}\simeq\gamma_{part}
\left(\frac{a_z}{a_{bb}}\right)^{1/2}
\label{partially0}
\end{equation}
where
\begin{equation}
\gamma_{part}=\left(c_1\frac{N_f^{1/2}}{N_b^{1/2}}+c_2\frac{N_b^{1/2}}{N_f^{1/2}}\right)^{-1}
\label{partially}
\end{equation}
with
\begin{equation}
c_1=\left(\frac{2}{\pi}\right)^{1/4}\!\!
\frac{m_f\omega_{xf}}{2m_r\omega_{xb}}
\left(\frac{\lambda_f}{\lambda_b}\right)^{1/2}
\end{equation}
and
\begin{equation}
c_2=\left(\frac{2}{\pi}\right)^{1/4}\!\!
\frac{m_b\omega_{xb}}{2m_r\omega_{xf}}
\left(\frac{\lambda_b}{\lambda_f}\right)^{1/2}.
\end{equation}
We recognise in Eq.~(\ref{partially}) a geometric combination of two
scaling parameters: $c_1{N_f^{1/2}}/{N_b^{1/2}}$
is dominant in the case $N_b\ll N_f$ while
$c_2{N_b^{1/2}}/{N_f^{1/2}}$ is dominant in the opposit limit.
The prediction obtained from Eq.~(\ref{partially0}) is shown in
Figs.~\ref{fig1} and \ref{fig2} by a solid arrow. 
There clearly is good
agreement  between the analytical estimate and the numerical results.

\subsection{Dynamical demixing}
On further increasing the boson-fermion coupling one reaches the point where
the fermion density vanishes at the centre of the trap. This occurs when
\begin{equation}
\left.\frac{g_{bf}}{g_{bb}}\right|_{dyn}=\frac{\mu_f}{\mu_b}.
\label{pablo}
\end{equation}
We denote this point as the dynamical location of the demixing in a 
mesoscopic cloud, since we expect a sharp upturn of the low-lying 
fermion-like collective mode frequencies to occur here as 
it is the case for both collisional and collisionless excitations 
in a mixture under 3D confinement ~\cite{Capuzzi2003a,Capuzzi2003b}.

If we insert the chemical potentials for ideal-gas clouds in 
Eq. (\ref{pablo}), we obtain an 
approximate expression for the location of dynamical demixing,
\begin{equation}
\left.\frac{a_{bf}}{a_{bb}}\right|_{dyn}\simeq\gamma_{dyn}
\left(\frac{a_z}{a_{bb}}\right)^{1/2}
\label{dyn0}
\end{equation}
where
\begin{equation}
\gamma_{dyn}=\left(\frac{\pi}{2}\right)^{1/4}\!\!
\frac{2m_f}{m_b+m_f}\frac{\omega_{xf}}{\omega_{xb}}
\left(\frac{N_f\lambda_f}{N_b\lambda_b}\right)^{1/2}.
\end{equation}
The condition for dynamical demixing coincides with that for
partial demixing in the limit $N_b\gg N_f$.
The prediction obtained from Eq.~(\ref{dyn0}) is indicated in
Figs.~\ref{fig1} and \ref{fig2} by a dot-dashed arrow.

\subsection{Full demixing}
\label{total_PS}
The point of full demixing is reached when the boson-fermion overlap 
becomes negligible as in a macroscopic cloud. Using the instability 
criterion outlined in Ref. \cite{Viverit2000a} for the 3D case, the 
condition for full phase separation at $T = 0$ is
\begin{equation}
\left.\frac{a_{bf}}{a_{bb}}\right|_{full}\simeq\gamma_{full}
\left(\frac{a_z}{a_{bb}}\right)^{1/2}
\label{condition}
\end{equation}
where
\begin{equation}
\gamma_{full}=\left(\sqrt{2\pi}\frac{2m_r}{m_b+m_f}\right)^{1/2}.
\end{equation}
At variance for the condition for full phase separation in 3D, for 
the Q2D confinement this criterion does not depend on the number of 
fermions, while $1/a_z$ 
plays the role of the Fermi momentum.

The prediction obtained from Eq. (\ref{partially0}) is indicated 
in Figs. \ref{fig1} and \ref{fig2} by a short-dashed arrow.

\subsection{Density profiles at full demixing}
In summary, it can be seen from 
Eqs.~(\ref{partially0}), (\ref{dyn0}) 
and (\ref{condition}) that the two control parameters for the 
transition in a mesoscopic cloud under Q2D confinement are
$a_{bf}/a_{bb}$ and $a_z/a_{bb}$.
Our main conclusion is that in a Q2D geometry a relatively 
small value of the boson-fermion scattering 
length suffices to reach the regime of full phase separation. 
In the case of the $^6$Li-$^7$Li mixture 
the bare values of the boson-boson and boson-fermion scattering lengths 
fulfill the condition for full phase separation for values 
of $a_z$ lower than 10 $a_{bf}$.

\begin{figure}[H]
\centerline{
\epsfig{file=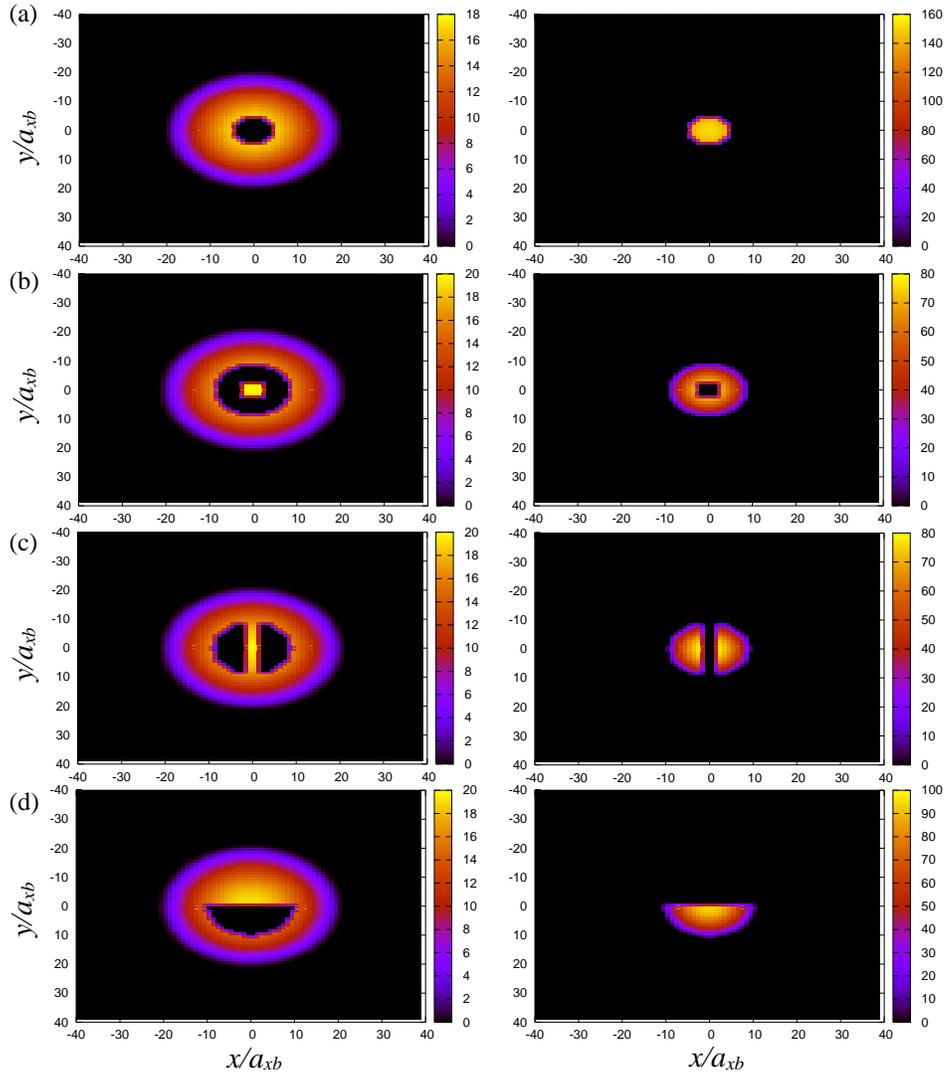,width=0.8\linewidth}}
\caption{Density profiles at full demixing for fermions (left) 
and bosons (right) in the $\{x,y\}$ 
plane, for a Q2D mixture with $a_{bb}=5.1\,a_0$, $a_{bf}= 38\,a_0$, and
$\lambda_b=\lambda_f=1$: 
(a) core of bosons surrounded by a ring of fermions; (b)
core of fermions surrounded by a ring of bosons inside a fermion 
cloud; (c) fermion slice between two boson slices 
surrounded by a fermion cloud; (d) core of bosons displaced away 
from the centre of the trap and surrounded by a fermion cloud.
Notice the difference in 
horizontal and vertical scales.}
\label{fig3}
\end{figure}

In view of the above result, it is useful to illustrate in 
Fig. \ref{fig3} the density profiles to be expected in the fully 
demixed regime for a mixture under isotropic confinement in the azimuthal 
plane (notice that a circular disc appears in Fig. \ref{fig3} as an oval, 
because of the different scales used on the horizontal and vertical axes). 
In fact
we have found various metastable configurations
for the demixed mesoscopic cloud in addition to 
the thermodynamically stable one consisting of a core of bosons 
surrounded by a ring of fermions.

In Fig. 3 we show topviews of the density profiles in the $\{x,y\}$ plane
for the choice of 
parameters $a_{bb}=5.1\,a_0$, $a_{bf}=38\,a_0$ 
and $a_z=a_{bf}$. 
The most energetically stable configuration is 
shown in Fig. \ref{fig3}(a) and lies at energy $E= 74.90\,N\hbar\omega_{xb}$.
Figure \ref{fig3}(b) shows a configuration with 
fermions at the centre surrounded by a ring of bosons inside a fermion 
cloud ($E = 78.94 \,N\hbar\omega_{xb}$). 
In Fig. \ref{fig3}(c) a fermion slice lies between two boson slices 
surrounded by a fermion cloud ($E = 79.04\,N\hbar\omega_{xb}$). 
Finally, Fig. \ref{fig3}(d) shows an asymmetric configuration in 
which a core of bosons is displaced away from the centre of the trap 
($E = 80.20\,N\hbar\omega_{xb}$). The latter two configurations
break the symmetry of the trap and this is an effect due to the finite
size of the system.

Configurations (b)-(d) are structurally 
but not energetically stable, {\it i.e.} they represent metastable
structures in local energy minima. They have been obtained by
using density profiles with various 
topologies as the initial guess for the self-consistent solution of
Eqs. (\ref{zehra1}) and (\ref{zehra4}).
In experiments bulk modes with the appropriate symmetry
may be exploited to attain these ``exotic'' 
configurations \cite{Akdeniz2003a}.

We have also studied the behaviour of the configurations shown in 
Fig. \ref{fig3} as a function of 
the scattering lengths ($5.1 a_0<a_{bb}<3\times 10^5 a_0$ and
$38 a_0<a_{bf}<5\times 10^7 a_0$) and the planar-anisotropy parameters
($10^{-3}<\lambda_{b,f}<400$). In contrast to the case 
of an elongated 3D confinement studied in our previous 
work \cite{Akdeniz2002a}, where the lowest-energy 
configuration can have different topologies depending on the system 
parameters, we have found that the configuration shown in 
Fig. \ref{fig3}(a) remains the most energetically stable for 
the Q2D geometry. A configuration consisting of a central core of 
fermions  surrounded by a cloud of bosons may, however, be the stable 
one for sufficiently strong boson-boson repulsions \cite{Akdeniz2002a}.

\section{Summary and concluding remarks}
\label{secconcl}
In summary, we have focussed on a boson-fermion mixture confined 
inside pancake-shaped traps, such that the scattering events can still 
be considered three-dimensional but nevertheless affected by the
vertical confinement. By using a 
mean-field description for the equilibrium densities in the 
azimuthal plane, we have studied the 
boson-fermion interaction energy as a function of the thickness 
of the atomic clouds and of the 
boson-boson and boson-fermion scattering lengths. 
We have given approximate analytical 
expressions identifying three critical regimes in terms of 
the scaling parameters $a_{bf}/a_{bb}$ and $a_z/a_{bb}$: 
(i) partial demixing where the boson-fermion interaction energy 
attains maximum value from a balance between increasing interactions 
and diminishing overlap; 
(ii) dynamical demixing where the fermionic density drops 
to zero at the centre of the trap and a sharp dynamical signature of demixing may be expected; and 
(iii) full demixing where the boson-fermion overlap is negligible
as in the macroscopic limit. 
These different criteria for the location of the phase transition 
in a mesoscopic cloud yield quite close predictions in a pancake-shaped 
trap under strong axial confinement.

We have found 
several metastable configurations for the demixed cloud, having various 
topologies but lying at higher energy above the stable configuration 
which is composed by a core of condensed bosons surrounded by fermions. 
We have verified that this is the minimum-energy configuration over 
extensive ranges of values for the coupling constants, the anisotropy 
in the plane of the trap, and the thickness of the trap.
	
A main result of our work is that full demixing can be reached 
in the $^6$Li-$^7$Li mixture by merely tuning the thickness of 
the trap, without necessarily tuning the scattering lengths. 
The full-demixing critical value of about $10\,a_{bf}$ for the thickness
is not attainable in the actual
experiments with the bare value of the $^6$Li-$^7$Li scattering length,
and a combination of squeezing of the pancake-shaped trap
and exploiting Feshbach resonances should be used.
On the contrary, the dynamical location of demixing may be observed
by only varying the number
of particles, without enhancing the scattering length.
In fact for a mixture of $10^6$ atoms of $^7$Li and $10^4$
atoms of $^6$Li trapped in the radial confinement discussed
in Sec. \ref{partial_sep}, it may be possible to observe a sharp
upturn of the low-lying fermionic modes for a pancake thickness 
of the order of $\sim 600\,a_{bf}\simeq 1\,\mu$m.

\vspace{1cm}

This work was partially supported by INFM under PRA-Photonmatter. 
ZA also acknowledges support from TUBITAK and from the Research Fund 
of the University of Istanbul under Project Number 161/15102003.


\end{document}